\documentclass{appolb}
\usepackage{epsfig}
\usepackage{wrapfig}
\usepackage{amssymb,amsmath} 

\newcommand {\pp} {\mbox{$p+p$}~}                                               
\newcommand {\ppbar} {\mbox{$p+\bar{p}$}~}                                               
\newcommand {\epem} {\mbox{$e^++e^-$}~}                                               
\newcommand {\AuAu} {\mbox{$Au+Au$}~}                                           
\newcommand {\rootsNN} {\mbox{$\sqrt{s_{NN}}$}}   
\newcommand {\KsKsC} {\mbox{$(K_s^0,K_s^0)$}}   
\newcommand {\ppC} {\mbox{$(p,p)$}}   
\newcommand {\pbarpbarC} {\mbox{$(p,\bar{p})$}}   
\newcommand {\llC} {\mbox{$(\Lambda,\Lambda)$}}   
\newcommand {\lbarlbarC} {\mbox{$(\bar{\Lambda},\bar{\Lambda})$}}   

\begin{document}
\title{Femtoscopy in hadron and lepton collisions: RHIC results and world systematics
\thanks{Presented at IV Workshop on Particle Correlations and Femtoscopy, Krak\'ow, Poland}%
}
\author{Zbigniew Chaj\c{e}cki
\address{Department of Physics, Ohio State University,\\
191 West Woodruff Ave, Columbus, OH 43210, USA}
}
\maketitle
\begin{abstract}
Femtoscopic measurements at a variety of facilities have established
a clear dependence of spatial scales with event multiplicity and particle transverse mass ($m_T$)
in heavy ion collisions from $\rootsNN\sim 2-200$~GeV.  The $m_T$-dependence is thought to arise
from collective, explosive flow of the system, as probed by independent measurements, while the
multiplicity dependence reflects the increased spatial extent of the final state with decreasing
impact parameter.
Qualitatively similar dependences have been reported from high energy hadron and lepton
collisions, where the conceptual validity of an impact parameter or collective flow are 
less clear.
We focus on results from elementary particle collisions, identify trends seen 
in the experimental data 
and compare them to those from heavy ion collisions.
\end{abstract}

\section{Introduction}

The enhancement of the probability of having two bosons close in phase-space is a consequence of Bose-Einstein
symmetrization. In astronomy, this effect was first observed by Hunbary Brown and Twiss~\cite{Hanbury:1954wr}
who measured the angular size of stars using photon intensity interference. 
For particle physicists, it all started about 50 years ago when Goldhaber {\it et al.}
observed significant positive correlations between identical pions due to 
the Bose-Einstein effect~\cite{Goldhaber:1960sf}. 
It was almost two more decades before the availability of data with sufficient statistics and quality
allowed pion interferometry to reliably extract spatial scales on the order of the interaction region, $\sim 1$~fm.
Later, with new facilities being built, the femtoscopic studies were extended to probe the sizes
of the particle emitting sources as a function of the initial system size and the energy of the
collisions
as well as  as a function of kinematic variables like the rapidity and the transverse momentum 
of the particle pair.  The interferometry method has been applied not 
just to pions but also to heavier particles.  
In principle, a systematic comparison of 
femtoscopic results from the elementary particle collisions (e.g. \pp, \ppbar, \epem) 
through light nuclei (e.g. $O+O$) and up to heavy ion collisions (e.g. $Au+Au$)
gives an opportunity to understand the physics of these collisions probed by two-particle 
interferometry.  In practice, such a comparison has been hampered by two problems.  Firstly,
there have traditionally been a multitude of parameterizations of the Bose-Einstein effect,
especially within the particle physics community.  This, coupled with different analysis techniques,
definitions of multiplicity, and acceptances, makes comparison of results between different experiments
difficult.  Secondly, communication between the high energy and heavy ion communities has unfortunately
been rather limited; workshops such as this one should help rectify this situation.

This article is not a complete review of femtoscopic results from high energy collisions.
Instead, we have collected a large fraction of the world dataset of such results and 
compare them in a common systematics.  We
focus on aspects of femtoscopy in elementary particle collisions that
are important when comparing to heavy ion collisions. We also discuss some issues 
with results from \epem collisions that complicate their interpretation. 

The paper is organized as follows.
In Section~\ref{sec:params} we present different parameterizations of one- and three-dimensional
correlation functions used to obtain results presented in this paper.
A short summary of the most important observables in femtoscopic studies in heavy ion
collisions is presented in Section~\ref{sec:femtobig}. In Section~\ref{sec:worldsystematics} we attempt 
to present a consistent picture of the femtoscopic observables in the elementary particle collisions.
In Section~\ref{sec:femtosmall} we bring up a problem with mass dependence of HBT radii seen in 
\epem collisions and comment on the Heisenberg uncertainty principle as a possible origin of both 
the mass ordering and the transverse mass dependence of $R_z$.
 Conclusions and discussion are presented in Section~\ref{sec:summary}.

\section{Definitions and parameterizations of Bose-Einstein effect}
\label{sec:params}

The correlation function is defined as
\begin{equation}
\label{eq:c2}
C(p_1,p_2) = \frac{P(p_1,p_2)}{P(p_1)P(p_2)},
\end{equation}
where $P(p_1,p_2)$ is the probability of observing two particles with momenta 
$p_1$ and $p_2$, while  $P(p_1)$ and $P(p_2)$ denote single-particle probabilities.

Experimentally, the correlation function is defined as 
\begin{equation}
\label{eq:CFexp}
C(Q)=\frac{A(Q)}{B(Q)},
\end{equation}
where $Q$ is a difference between momenta of two particles.
$A(Q)$ represents a distribution of the pairs from the same event and $B(Q)$ 
is the background, or reference, distribution that is supposed to include all physics 
effects as $A(Q)$ except for femtoscopic correlations
(quantum statistics, final state interactions including Coulomb and strong interaction, 
where applicable)  between the pair being studied.

Femtoscopic correlations in $\vec{Q}$-space may be expressed (e.g~\cite{Lisa:2005dd})
as a convolution of the pair spatial separation distribution with the two-particle
wavefunction which includes quantum (anti-)symmetrization and final state interactions effects.
At large $|\vec{Q}|$, these effects vanish, and femtoscopic correlation functions must assume
a constant value independent of the direction of $\vec{Q}$.
However, several experiments of collisions with low 
multiplicity (e.g.~\cite{Avery:1985qb,Agababyan:1996rg,collaboration:2007he,Bailly:1988zb,Uribe:1993tr})
report correlation functions with large-$|\vec{Q}|$ structure which must be non-femtoscopic in origin.  They
may arise, for example, from jets or energy-momentum conservation.

Especially in \epem experiments, there has been tremendous effort to remove these effects using
different techniques to form the reference distribution, $B(q)$ in
Eq.~\ref{eq:CFexp} (e.g.~\cite{Alexander:2003ug}).
For identical particle interferometry, these include using like-sign pion distributions or Monte-Carlo simulations
to generate the reference.  We do not review them all here, simply noting that
each technique has its advantages, but none completely removes non-femtoscopic
structures.  At the end, the correlation function is fit with a functional form which includes
(or neglects) the non-femtoscopic structure.  A plethora of forms has been used over the years in
high-energy particle measurements.  We discuss some here.

Usually, one assumes that the measured correlation function approximately factors into 
a femtoscopic ($C_F\left(\vec{q}\right)$) and a non-femtoscopic  ($\zeta\left(\vec{q}\right)$) part.
\begin{equation}
\label{eq:factorization}
C\left(\vec{q}\right) = C_F\left(\vec{q}\right) \cdot \zeta\left(\vec{q}\right) .
\end{equation}

\subsection{Femtoscopic forms}
\label{sec:femto}

Here, we list some fitting forms used in the high energy literature.  If non-femtoscopic effects are ignored
($\zeta=1$), then these are the forms used to fit the measured correlation function.

The 1D HBT radius can be obtained by fitting the correlation function (Eq.~\ref{eq:CFexp})
with an analytical parameterization.
The most commonly used one, that assumes the Gaussian source distribution, is defined as 
\begin{equation}
\label{eq:CFqinv}
C_F(Q_{inv})=  1 + \lambda e^{-Q^2_{inv} R^{2}_{inv}},
\end{equation}
where $Q_{inv} \equiv \sqrt{(\vec{p_1} - \vec{p_2})^2 - (E_1 - E_2)^2}$.

Another parameterization, also assuming a Gaussian shape of the source distribution,
 but this with time $Q$ is measured in the lab frame, is
\begin{equation}
\label{eq:CFgauss}
C_F(q,q_0)=  1 + \lambda e^{-q^2 R^{2}_{G} - q^2_0 \tau^2 },
\end{equation}
where $q=|p_1-p_2|$, $q_0=E_1-E_2$, $R_{G}$, $\tau$ and $\lambda$ are the source size, lifetime and chaoticity parameter.

Kopylov and Podgoretskii~\cite{Kopylov:1972qw} introduced an alternative parameterization
\begin{equation}
\label{eq:CFcp}
C_F(q_T,q_0)=  1 + \lambda \left[ \frac{2 J_1 \left( q_T R_B \right) }{q_T R_B}\right]^2 \left(1 +  q^2_0 \tau^2 \right)^{-1},
\end{equation}
where $q_T$ is the transverse component of $\vec{q}=\vec{p_1}-\vec{p_2}$ with respect to $\vec{p}=\vec{p_1}+\vec{p_2}$, $q_0=E_1-E_2$, $R_B$ and $\tau$ are the size and decay constants of
a spherical emitting source,
and $J_1$ is the first order Bessel function.  

Simple numerical studies 
show that $R_{G}$ from 
Eq.~\ref{eq:CFgauss} is approximately twice smaller than $R_{B}$ obtained from Eq.~\ref{eq:CFcp} (e.g.~\cite{Boal:1990yh,Alexopoulos:1992iv}).

With enough statistics, a femtoscopic analysis may be performed in 2 or 3 dimensions.
Then, the correlation function 
is often expressed in the Bertsch-Pratt decomposition~\cite{Bertsch:1988db,Pratt:1990zq} 
\begin{equation}
\label{eq:cf3D}
C_F(q_o,q_s,q_l)=  1 + \lambda e^{-q^2_o R^2_o - q^2_s R^2_s - q^2_l R^2_l},
\end{equation}
where, $\vec{Q}=(q_o,q_s,q_l)$ is defined in the longitudinally co-moving frame, $q_l$ is the component
parallel to the beam axis or trust axis (in \epem), $q_o$ is measured in transverse plane and points 
into the direction of outgoing pair and $q_s$ is perpendicular to other two components.
Analogously, the sizes of the source along these three directions are denoted as $R_o$, $R_s$ and $R_l$.

A similar form is used for two-dimensional correlation functions:
\begin{equation}
\label{eq:cf2D}
C_F(q_T,q_l)=  1 + \lambda e^{-q^2_T R^2_T - q^2_l R^2_l},
\end{equation}
where, $q_T=\sqrt{q_o^2 + q_s^2}$.

Other groups have used double-Gaussian~\cite{Akesson:1987bz,Eggers:2006mh} and exponential~\cite{Adloff:1997ea} fitting functions.  A scan of
the literature reveals even more.
We do not wish to tabulate all the myriad forms used in high energy physics, but simply point out that this
lack of a consistent analysis is rather a plague, greatly impeding any effort to make sense of the program
as a whole.

\subsection{Non-femtoscopic forms}
\label{sec:nonfemto}

Many experiments ignore any non-femtoscopic contributions to the measured correlation function, setting
$\zeta=1$ (c.f. Equation~\ref{eq:factorization}) and
simply fitting it with one of the forms above.  Of those that account for the long-range structure, most
simply choose a functional form that seems to describe the correlation at large $Q_{inv}$, $q$ or $q_T$, depending on the coordinate system used in the analysis, {\it assume} that
it extrapolates smoothly into the femtoscopic region, and fit.  These ad-hoc forms have no real foundation
in physics, but are simply guesses.

The most popular form, used e.g. in~\cite{Akers:1995ay,Abbiendi:2000gc,Abbiendi:2000jb,Abbiendi:2003fq,Schael:2004qn}, is
\begin{equation}
\zeta\left( Q_{inv}\right) = 1 + \epsilon\cdot Q_{inv} + \delta\cdot Q_{inv}^2 ,
\end{equation}
where $\epsilon$ and $\delta$ are free parameters.
In same cases, this form was modified by setting $\delta=0$~\cite{Buskulic:1994ny,Achard:2001za,AguilarBenitez:1991ri,Abreu:1994wd,Abreu:1992gj} or $\epsilon=0$~\cite{AguilarBenitez:1991ri}.

An alternative one-parameter form was introduced in~\cite{Bailly:1988zb}
\begin{equation}
\zeta\left( Q_{inv}\right) = \left(1 + \delta\cdot Q_{inv}^2 \right)^{-1} .
\end{equation}
For simplicity, we expressed above forms as a function of $Q_{inv}$ only, however, 
one can easily replace $Q_{inv}$ with e.g. $q$ or $q_T$.

While expressing it analytically here would require additional explanation,
yet another ad-hoc form was introduced by STAR~\cite{Chajecki:2005iv}
to describe the three-dimensional structure of the
correlation function with two additional parameters.

Finally, a formulation has recently been proposed~\cite{Chajecki:2008vg} which is not ad-hoc,
but provides a full three-dimensional analytic form for $\zeta$, assuming that the non-femtoscopic 
correlations arise from energy and momentum conservation.  Recording the formula here would require
extensive explanation, so we refer the reader to~\cite{Chajecki:2008vg} for details.  In this approach,
the extra ``parameters'' are physical quantities like the total multiplicity, average energy, and so forth.

There are still other ad hoc forms not listed here.  The reader should be impressed
(or depressed) by the potential combinatorics of combining
the $\zeta$ terms from this Section with the $C_F$'s from Section~\ref{sec:femto}.

\section{Two Scalings in Femtoscopic Results in Heavy Ion Collisions}
\label{sec:femtobig}

A systematic overview of femtoscopic studies in relativistic heavy ion collisions reveals a broad variety
of interesting trends reflecting the underlying system dynamics~\cite{Lisa:2005dd}.  Here, we mention two
of the most important, shown in Figure~\ref{fig:scalings}.


\subsection{Transverse mass dependence}
\begin{figure}[ht]
  \begin{center}
\includegraphics[width=0.45\textwidth]{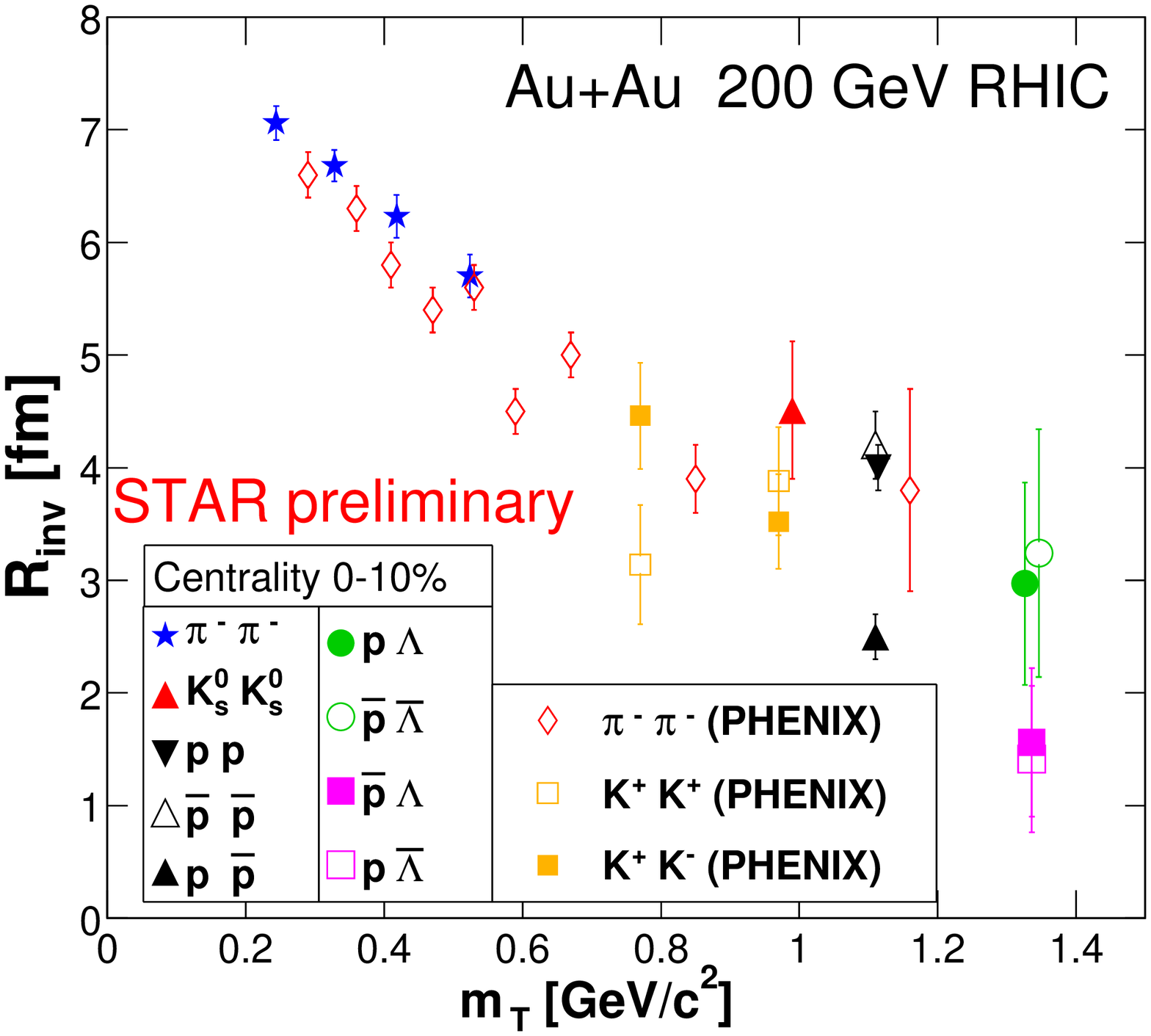}
\includegraphics[width=0.35\textwidth]{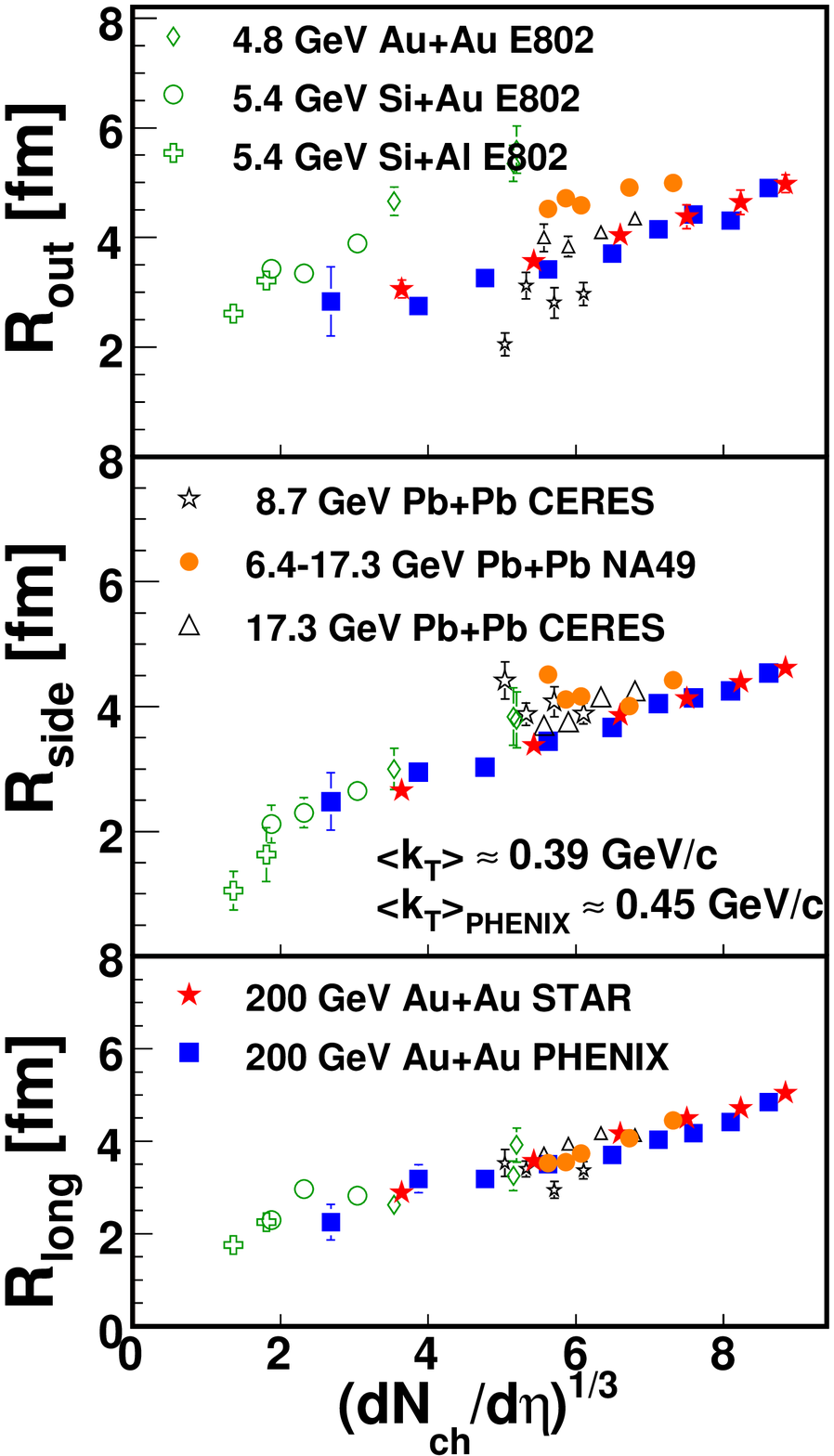}
\end{center}
\caption{Left: $m_T$ dependence of $R_{inv}$ for different particles. Figure taken from~\cite{Chajecki:2005iv}.
Right: Femtoscopic radii dependence on the number of charged particle. Figure taken from~\cite{Lisa:2008gf}. }
\label{fig:scalings}
\end{figure}
 The negative correlation between the femtoscopic sizes and the transverse mass of the particles
is usually attributed to collective flow of a bulk system~\cite{Pratt:1984su}.
In such a scenario, 
 approximately ``universal'' $m_T$ dependence of femtoscopic radii applies not only to pions but to all
particle types.  This is in fact observed experimentally and results 
have been presented in the left panel of Figure~\ref{fig:scalings}, in which 
one-dimensional radii from pion~\cite{Heffner:2004js}, charged kaon~\cite{Heffner:2004js},
 neutral kaon~\cite{Bekele:2004ci}, 
proton and anti-proton~\cite{Gos:2006fe}, and proton-$\Lambda$~\cite{Chaloupka:2005hj} correlations are plotted.
Characteristic signals of collective flow~\cite{Adams:2003qa} are also observed in 
correlations between particles with very different masses.


\subsection{Multiplicity scaling}
The right panel of Figure~\ref{fig:scalings} presents  AGS/SPS/RHIC systematics of HBT radii 
dependence on $(dN_{ch}/d\eta)^{1/3}$ ($N_{ch}$ - number of charged particles) for different colliding systems 
at different energies of the collisions. The main motivation for studying such  
a relation is its connection to the final state geometry 
through the particle density at freeze-out.  
As seen, all radii exhibit a scaling with $(dN_{ch}/d\eta)^{1/3}$.
It is especially interesting that the radius parameters $R_{side}$ and 
$R_{long}$ follow the same trend for different collisions over a wide range 
of energies and given value of $\langle k_T\rangle$.
It is a  clear signature 
that the multiplicity is a scaling variable that drives these geometrical radius parameters. 
Since $R_{out}$ mixes space and time information it is not clear
whether one expects its simple scaling with the final state geometry~\cite{Lisa:2005dd}.

\section{World systematics from elementary particle collisions}
\label{sec:worldsystematics}
\begin{table}
\begin{tabular}{|c|c|c|c|c|}
\hline
System & $\sqrt{s} [GeV]$ & Facility & Experiment & Refs. \\
\hline
\hline
\pp & ~1.9 & LEAR  &  CPLEAR & \cite{Adler:1994gr,Angelopoulos:1997ix} \\
   & ~1.9 & CERN &  ABBCCLVW & \cite{Deutschmann:1982vz} \\
    & 7.2 & AGS & E766 & \cite{Uribe:1993tr} \\
    & 17  & SPS  & NA49  & \cite{Ganz:1999ht}\\
    & 26  & SPS  & NA23 & \cite{Bailly:1988zb} \\
    & 27.4 &SPS  & NA27 & \cite{AguilarBenitez:1991ri} \\    
    & 31-62 & ISR  & AFS & \cite{Akesson:1985ax,Akesson:1986ix,Akesson:1987bz} \\
    & 44,62 & ISR & ABCDHW & \cite{Breakstone:1986xs} \\
    & 200 & SPS & NA5 & \cite{DeMarzo:1984zr} \\
    & 200 & RHIC & STAR &  \cite{Chajecki:2005zw}\\
\hline
\ppbar  & 53 & ISR & AFS & \cite{Akesson:1983gk} \\
    & 200  & SPS  & NA5 &\cite{DeMarzo:1984zr} \\
    & 200-900  & SPS & UA1 & \cite{Albajar:1989sj} \\
    & 1800  & Tevatron & E735  & \cite{Alexopoulos:1992iv} \\
\hline
$h+p$ & 5.6 & CERN &  ABBCCLVW & \cite{Deutschmann:1982vz} \\
      & 21.7  & SPS  & EHS/NA22 & \cite{Agababyan:1993aj,Agababyan:1996rg} \\
\hline
$e^{+}+e^{-}$ & 3-7,29  & SLAC  & Mark-II & \cite{Juricic:1988zd} \\
  & 10  & CESR  & CLEO  & \cite{Avery:1985qb}\\
  & 29  & SLAC  & TPC  & \cite{Aihara:1984mm} \\
  & 29-37  & DESY-PETRA  & TASSO  & \cite{Althoff:1985ws,Althoff:1986wn} \\
  & 58  & TRISTAN  & AMY  & \cite{Choi:1995xb}\\
  & 91  & LEP  & OPAL & \cite{collaboration:2007he,Abbiendi:2003fq,Alexander:1996jq,Abbiendi:2000jb,Akers:1995ay,Achard:2001za} \\
  & 91  & LEP  & L3  & \cite{Achard:2001za}\\
  & 91  & LEP  & DELPHI  & \cite{Smirnova:1999aa,Abreu:1992gj,Abreu:1994wd,Abreu:1996hu,Abreu:1994we,Delphilambdalambda:2004}\\
  & 91  & LEP  & ALEPH  & \cite{Decamp:1991md,Heister:2003ai,Schael:2004qn,Barate:1999nv} \\
\hline
$e^{+}p$  & 300  & HERA  & ZEUS  & \cite{Chekanov:2003gf,Chekanov:2007ev}\\
   &  300  & HERA  & H1  & \cite{Adloff:1997ea}\\
\hline
$\mu p$ & 23 & CERN & EMC-NA9 & \cite{Arneodo:1986bza} \\
\hline
$\alpha+\alpha$    & 126  & ISR  &  AFS & \cite{Akesson:1983gk,Akesson:1985ax,Akesson:1986ix,Akesson:1987bz} \\
\hline
$\mu N$ & 30 & Tevatron & E665 & \cite{Adams:1993hd} \\
\hline
$\nu N$ & $>$10 & & BBNC & \cite{Korotkov:1993my} \\
\hline
\end{tabular}
\caption{ Collection of published experimental studies of two-particle correlations in small systems.
\label{tab:ta}}
\end{table}
The Bose-Einstein effect has been studied in elementary particle collisions 
for decades by various experiments.
We made attempt to collect
experimental papers on two-particle correlations in small systems  in Table~\ref{tab:ta}.
Here, we will focus on results from elementary particle collisions by looking 
at the multiplicity and transverse mass dependence of femtoscopic sizes measured in these
collisions and we will compare them to those from heavy ion collisions (discussed in
Section~\ref{sec:femtobig}).
\subsection{Multiplicity dependence}
\label{sec:multdepsmall}
\begin{figure}[t!]
{
\centerline{\includegraphics[width=0.78\textwidth]{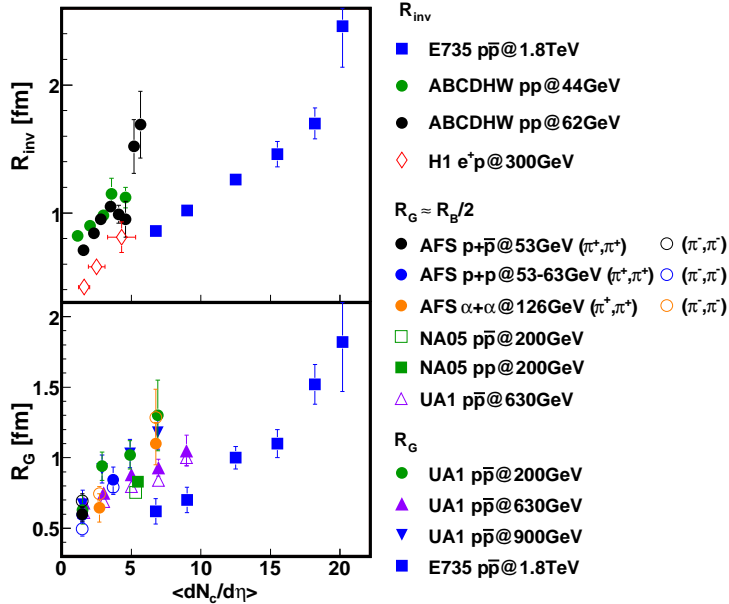}}}
\caption{The multiplicity dependence of the pion HBT radii.Compilation of results from various experiments. Only data from collisions at $\sqrt{s} > 40~GeV$ are shown.  }
\label{fig:MultDep1D}
\end{figure}
\begin{figure}[t]
{\centerline{\includegraphics[width=0.99\textwidth]{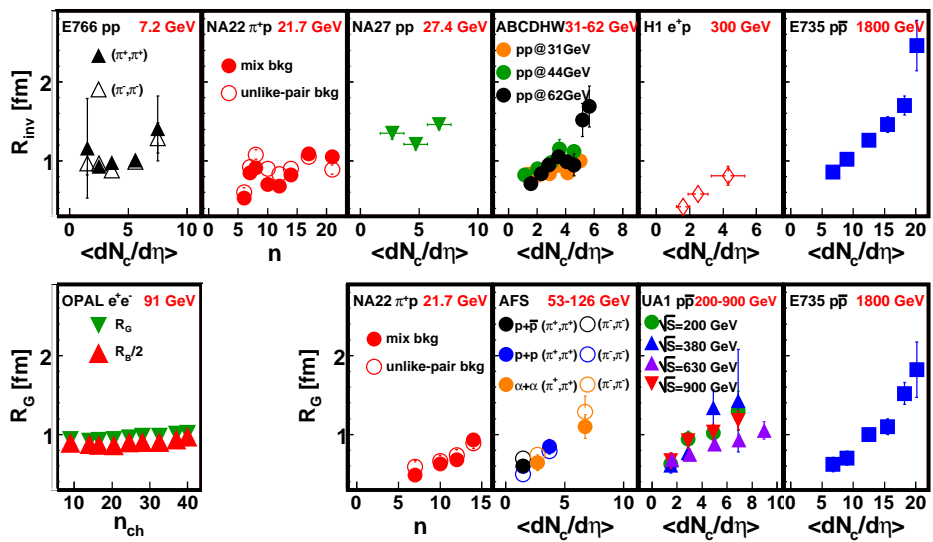}}}
\caption{Multiplicity dependence of pion HBT radii presented as a function of the energy of the 
collisions from the following experiments
 E735~\cite{Alexopoulos:1992iv}, 
 ABCDHW~\cite{Breakstone:1986xs},
 UA1~\cite{Albajar:1989sj},
 AFS~\cite{Akesson:1983gk},
 NA5~\cite{DeMarzo:1984zr},
 E766~\cite{Uribe:1993tr},
 NA22~\cite{Agababyan:1993aj},
 NA27~\cite{AguilarBenitez:1991ri}.
 H1~\cite{Adloff:1997ea},
 OPAL~\cite{Alexander:1996hb}.
}
\label{fig:MultDep1DSep}
\end{figure}

Figure~\ref{fig:MultDep1D} shows a collection of results from a number of experiments studying 
\pp, \ppbar, \epem and even $\alpha-\alpha$ collisions plotted versus the number of charged 
particles per unit of pseudorapidity. $R_{inv}$ (c.f. Eq.~\ref{eq:CFqinv}) is plotted on upper panel and 
$R_G$ (c.f. Eq.~\ref{eq:CFgauss}) on the lower panel of this figure. 
Since $R_B \approx 2 R_G$
we were able to plot both radii together  by dividing $R_B$ (c.f. Eq.~\ref{eq:CFcp}) radii by a factor of 2.
All radii shown are from collisions with $\sqrt{s}>40$~GeV and increase with multiplicity.
For collisions with $\sqrt{s}<40$~GeV, the multiplicity systematics are less clear, since some
experiments report a clear relationship and some do not, as shown in detail in Figure~\ref{fig:MultDep1DSep}.
While the multiplicity dependences from different experiments seen in Figure~\ref{fig:MultDep1D}
are qualitatively similar, we do not see the quantitatively universal dependence observed in heavy
ion collisions (c.f. right side of Fig.~\ref{fig:scalings}).
This may indicate that, in fact, there is no such universal scaling in particle collisions.
However, even if there were a universal multiplicity dependence of femtoscopic scales,
there are at least three other reasons for which such a scaling would not appear in Figure~\ref{fig:MultDep1D}.

Firstly, as discussed in Section~\ref{sec:params}, the various experiments use different fitting functions
to extract radius parameters.  Even when extracting the seemingly straightforward parameter $R_{inv}$ with
the form given by Equation~\ref{eq:CFqinv}, different parameterizations of the non-femtoscopic term $\zeta$
were used to fit the measured correlation function; c.f. Equation~\ref{eq:factorization}.

Secondly, as discussed below, the femtoscopic scales depend not only on multiplicity, but also
kinematic quantities such as transverse momentum $p_T$.  The various experiments had significantly
different acceptances, for which it is hard to account after the fact; this will lead to systematic
differences between one experiment's results and another's.  For example, the Tevatron experiment
E735~\cite{Alexopoulos:1992iv} was more biased towards high-$p_T$ particles than the other experiments.
Thus, one expects E735's radii to be systematically lower (for fixed multiplicity) than the others;
this is precisely what is seen in Figure~\ref{fig:MultDep1D}.

Thirdly, these experiments often use quite different definitions of ``multiplicity'' in their publications,
making apples-to-apples comparisons more difficult.  As best we could, we tried to compare results as a function
of $dN_{ch}/d\eta$.
For example, sometimes, the number of all particles per unit rapidity is reported; in such case we assumed
that charged particles are two-third of the total multiplicity. 
Some experiments provide the number
of charged particles in some range of pseudorapidity (not always centered at $\eta=0$).
In such cases we assumed a flat $\eta$ distribution, and scaled.
Additionally, we may expect some unknown bias between experiments in the method of extracting 
the number of (charged) particles, since these multiplicities were often ``raw'' numbers, not corrected
for efficiency.

In general, one would like to assign systematic errors from all of these effects, to see whether 
the data are indeed consistent with a ``universal'' multiplicity dependence.  However, beyond
what is described above, we were unable to do this from the information published by the collaborations,
so settle for a study of trends.
We note that these complicating issues are not problems for the heavy ion data, in which the
communication between different experiments is good, and apples-to-apples comparisons are seen as a high priority.

\subsection{The transverse mass/momentum dependence}
\label{sec:smallptdep}
\begin{figure}[ht]
{\centerline{
\includegraphics[width=0.45\textwidth]{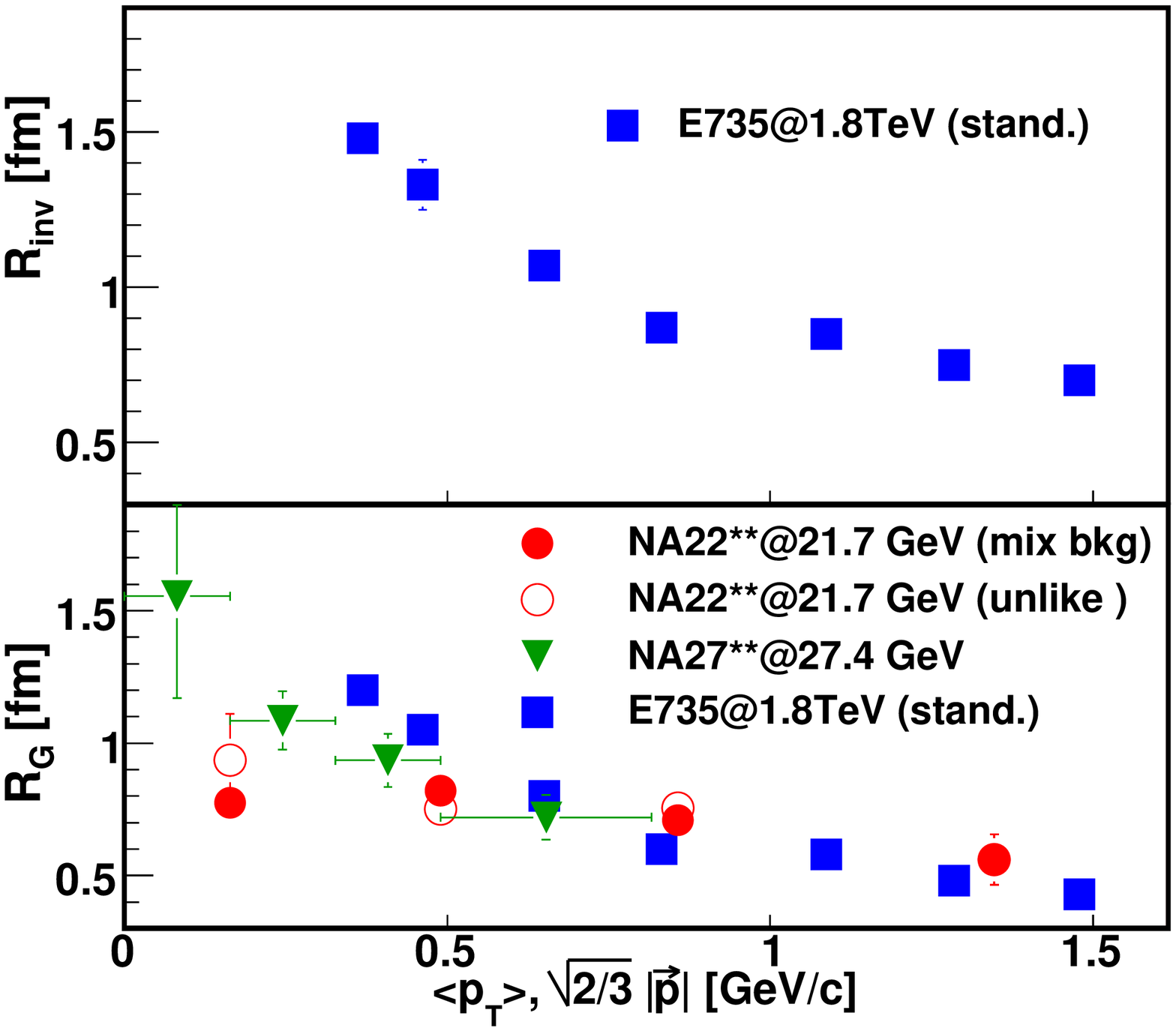}
\includegraphics[width=0.45\textwidth]{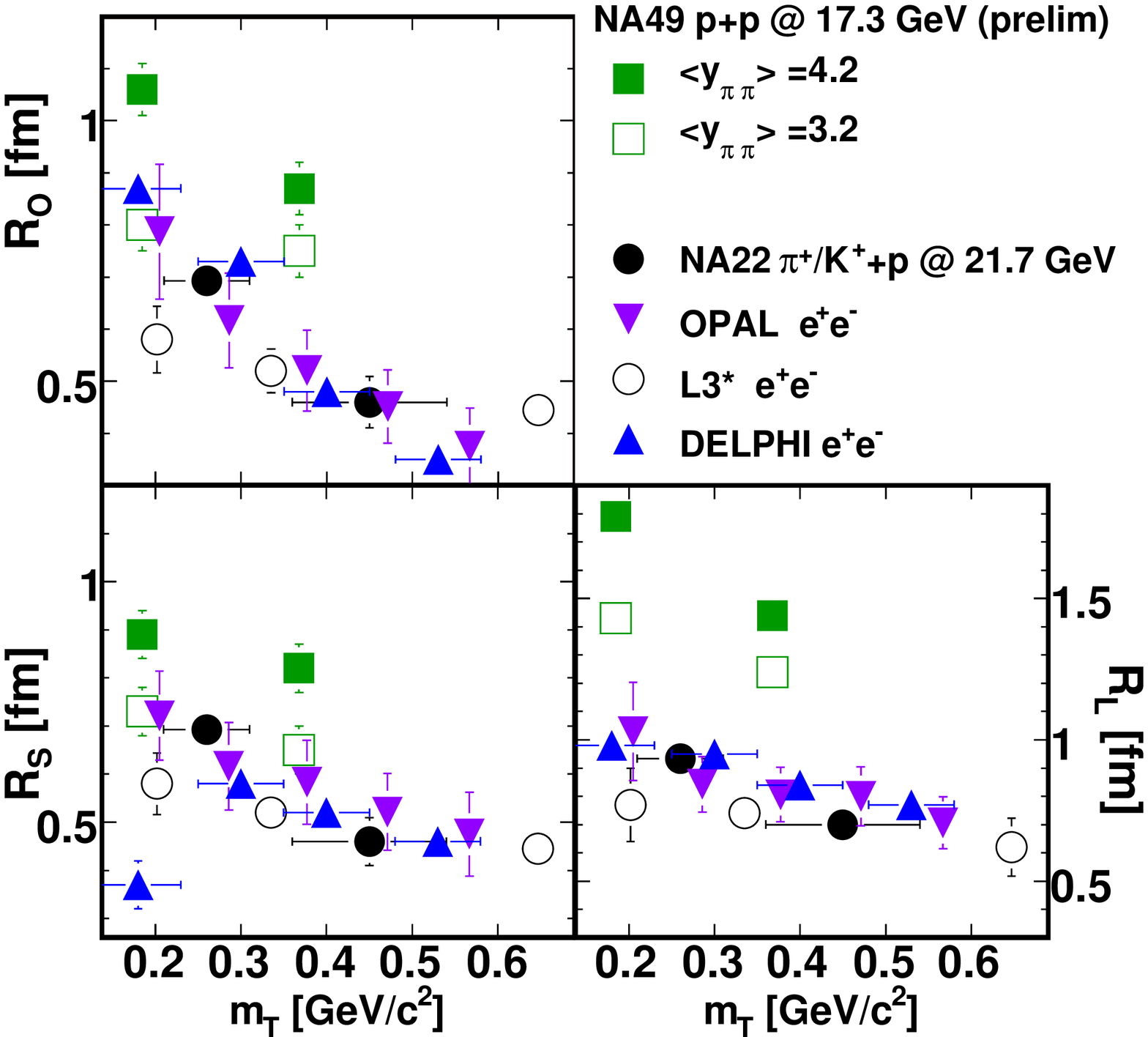}
}}
\caption{The transverse mass dependence of $R_{inv}$ (left panel) and 
the 3D HBT radii (right panel) from elementary particle collisions. Data from
 NA22~\cite{Agababyan:1996rg},
 NA49 preliminary~\cite{Ganz:1999ht},
 OPAL~\cite{collaboration:2007he},
 L3~\cite{Achard:2001za},
 DELPHI~\cite{Smirnova:1999aa}.
}
\label{fig:mTdep}
\end{figure}
In heavy ion collisions, the observed decrease of the HBT radii with increasing transverse mass 
has been associated with flow, as mentioned in Section~\ref{sec:femtobig}.
Whether or not it arises from the same physics, a similar $m_T$ scaling is observed in small systems,
as seen in Figure~\ref{fig:mTdep}.
Recalling the discussion from Section~\ref{sec:multdepsmall}, we recall that these results
come from different experiments, and that high energy particle measurements are hard to compare
quantitatively.  In particular,
the average number of particles per unit of pseudorapidity is different 
in each experiment and as discussed above, the magnitude of the HBT radii
depends on this value.  Additionally, as also discussed previously, 
somewhat different functional forms were used to extract the radii.
However, despite these difficulties it is clear that pion HBT radii in elementary particle collisions 
show similar dependence on the transverse momentum as seen in heavy ion collisions, though it is not
clear whether they have the same origin.

\section{Some aspects of femtoscopy in elementary particle collisions}
\label{sec:femtosmall}

In this Section, we will discuss two issues with femtoscopic results from
elementary particle collisions. The first one is related to experimental data and 
the second one to the interpretation of these data.

\subsection{Is there a mass dependence of HBT radii in \epem collisions?}
\label{sec:massdep}

As we showed in previous Section, the pion HBT radii from small systems 
are decreasing with increasing transverse {\it momentum}, similar with the trend from heavy ion collisions.
However, in heavy ion collisions, a scaling with transverse {\it mass} is seen for several particle types,
not only pions.
Thus, in this Section, we would like to investigate whether there is mass dependence
of femtoscopic radii in elementary particle collisions.
As there is insufficient data available from hadron-hadron collisions
we will focus on data from \epem collisions only.
There have been claims (e.g.~\cite{Alexander:1999rm}) of the mass ordering in these collisions,
and these claims have been so often repeated so as to become ``common knowledge.''
However, we argue that a survey of \epem publications reveals a much less clear picture.
\begin{figure}[hbtp]
  \begin{center}
\includegraphics[width=0.95\textwidth]{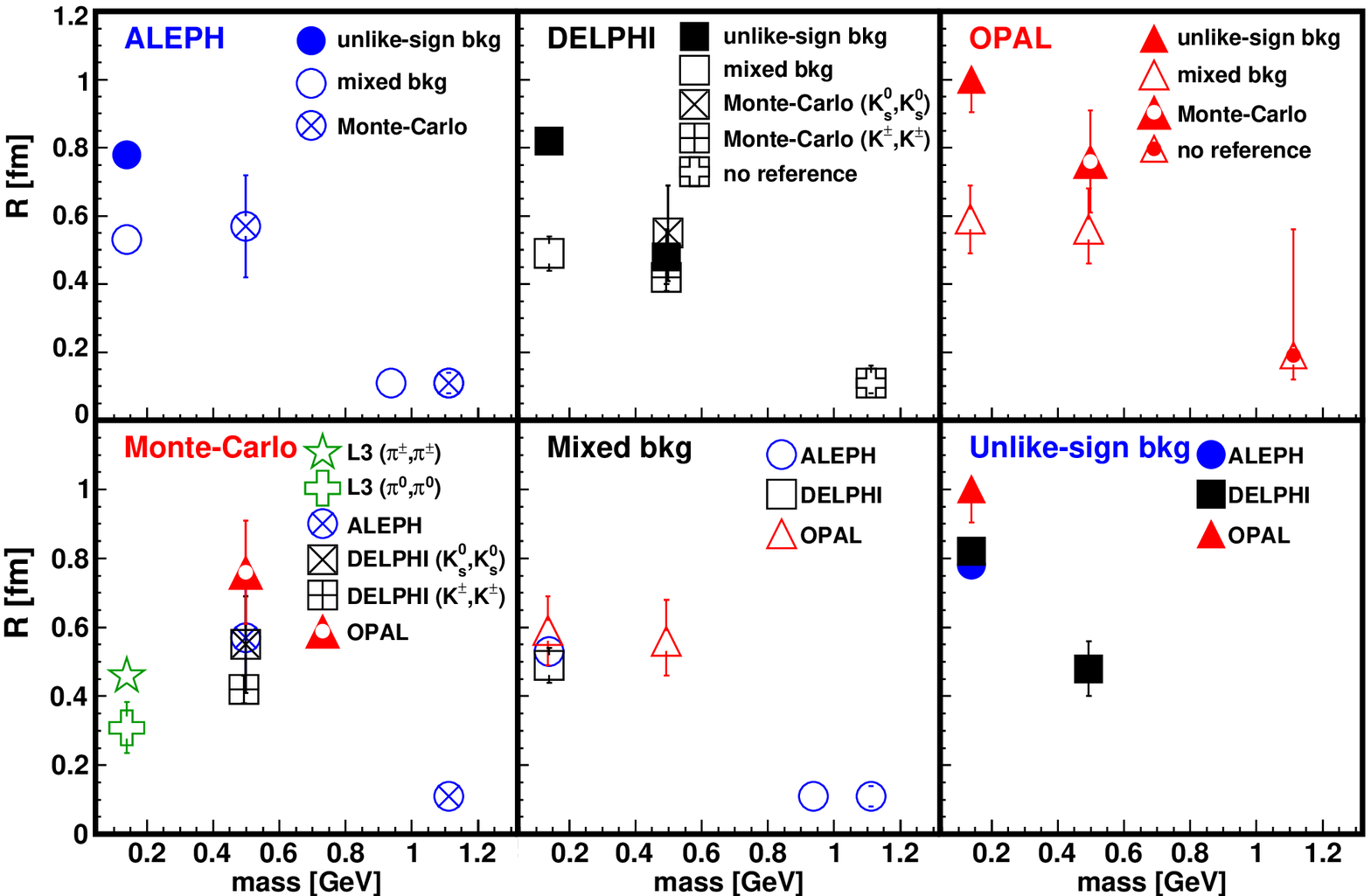}
    \caption{Results from \epem collisions. Data taken from the following publications: 
ALEPH - charged pions~\cite{Heister:2003ai},
\KsKsC, \ppC and \pbarpbarC~\cite{Schael:2004qn}, \llC and \lbarlbarC~\cite{Barate:1999nv}; 
DELPHI - charged pions~\cite{Abreu:1992gj,Abreu:1994wd}, charged kaons~\cite{Abreu:1996hu},
\KsKsC~\cite{Abreu:1994we} and \llC~\cite{Delphilambdalambda:2004};
OPAL - neutral pions~\cite{Abbiendi:2003fq}, \llC~\cite{Alexander:1996jq}, 
charged kaons~\cite{Abbiendi:2000jb}, \KsKsC~\cite{Akers:1995ay} and L3~\cite{Achard:2001za}. 
}
    \label{fig:epem}
  \end{center}
\end{figure}
We found two main reasons that complicate the systematic comparison of the data.
The first is the tendency of \epem experiments to use very different fitting forms for the
correlation function, even within one collaboration.  We have discussed this extensively above
in Sections~\ref{sec:params} and~\ref{sec:worldsystematics}, so only mention it here.
The second complication comes from the fact that experiments used various techniques 
to construct the reference distribution ($B(q)$ from Eq.~\ref{eq:CFexp}).

Results from four \epem experiments,
 ALEPH, DELPHI, OPAL and L3, are presented on Fig.~\ref{fig:epem}.
The upper panels show the mass dependence of HBT sizes measured using different 
techniques to construct the reference distribution from three experiments separately.
The bottom panels present the same data but this time grouped by the technique
of creating background of the correlation function. 
We make two interesting observations.
Firstly, not a single experiment used the same method to construct the  reference distribution
for pion, kaon and proton correlation function. Instead, they used different techniques for different
particles, and when we consider systematic errors that each technique introduces it is hard to make
any quantitative statement about the mass dependence of HBT radii. 
Secondly, when we group the data by a given method of reference generation (lower panels of Figure~\ref{fig:epem}),
identifying clear trends becomes more difficult.

We believe that the only fair statement that we can make about the femtoscopic results
from \epem collisions is that the size of the source for mesons is 
larger than that for baryons.  As Alexander pointed out~\cite{Alexander:2003ca}, this alone is a big problem for the Lund string model.

\subsection{Heisenberg uncertainty principle}
\begin{figure}[t]
  \begin{center}
\includegraphics[width=0.7\textwidth]{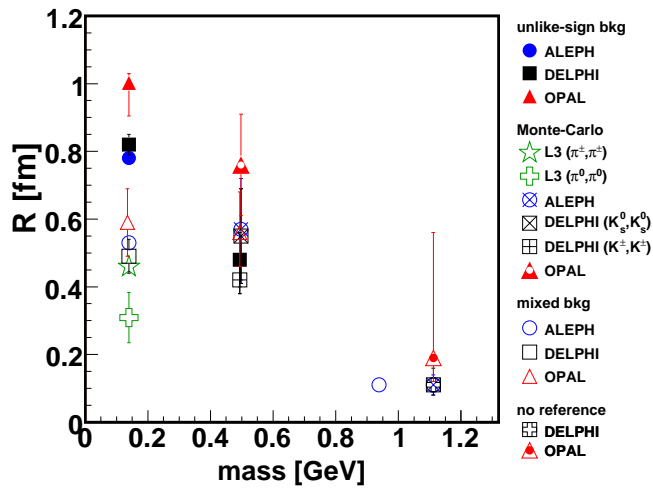}
    \caption{Combined results from \epem collisions. The same data as plotted on Fig.~\ref{fig:epem}.}
    \label{fig:allexps}
  \end{center}
\end{figure}
The Heisenberg uncertainty principle was pointed by Alexander {\it et al.}~\cite{Alexander:1999rm}
as a possible origin of the mass dependence of the HBT radii in \epem collisions.
Authors argue that the dependence of the one-dimensional radius $R_{inv}$ on the hadron mass obtained at LEP 
is consistent with the formula derived from the Heisenberg uncertainty principle 
\begin{equation}
\label{eq:HUR}
R_{inv}(m) = \frac{c \sqrt{\hbar \Delta t}}{\sqrt{m}}, 
\end{equation}
where $\Delta t$ was chosen to be $10^{-24}s~(0.3~fm)$.

However, the good agreement between the experimental data and the theoretical 
curve (Eq.~\ref{eq:HUR}) presented by the authors~\cite{Alexander:1999rm} raises 
a concern especially in the light of what we just discussed in the previous section.
In fact, when  all results from \epem experiments are plotted together 
as  done on Figure~\ref{fig:allexps}, we see that 
that it is difficult to make any quantitative statement about the mass ordering of HBT radii
due to a significant systematic bias on the radii coming from the different techniques to construct
the background of the correlation function and the spread of results coming from different experiments.

The one-dimensional radius provides limited information about the source that is 
in fact a three-dimensional distribution. Thus, to verify whether the Heisenberg uncertainty principle
can really explain the $m_T$ dependence of HBT radii one should look at three (eventually two) dimensional
radii. 
The DELPHI experiment~\cite{Smirnova:1999aa} report a decrease of three-dimensional radii with $m_T$;
c.f. Section~\ref{sec:worldsystematics}.
Alexander argues~\cite{Alexander:2001gk} that the Heisenberg uncertainty principle can best be  
applied to the longitudinal component of the radius $r_z$ (sometimes noted as $R_{L}$)
measured in LCMS frame using similar procedure as  it was done for $R_{inv}$\cite{Alexander:1999rm}.
The final formula gives approximately the same  dependence of the longitudinal femtoscopic size on 
$m_T$ as in the 1D case
\begin{equation}
\label{eq:HURRz}
R_{L}(m_T) \approx \frac{c \sqrt{\hbar \Delta t}}{\sqrt{m_T}}.
\end{equation}
\begin{figure}[t]
{
\centerline{
\includegraphics[width=0.5\textwidth]{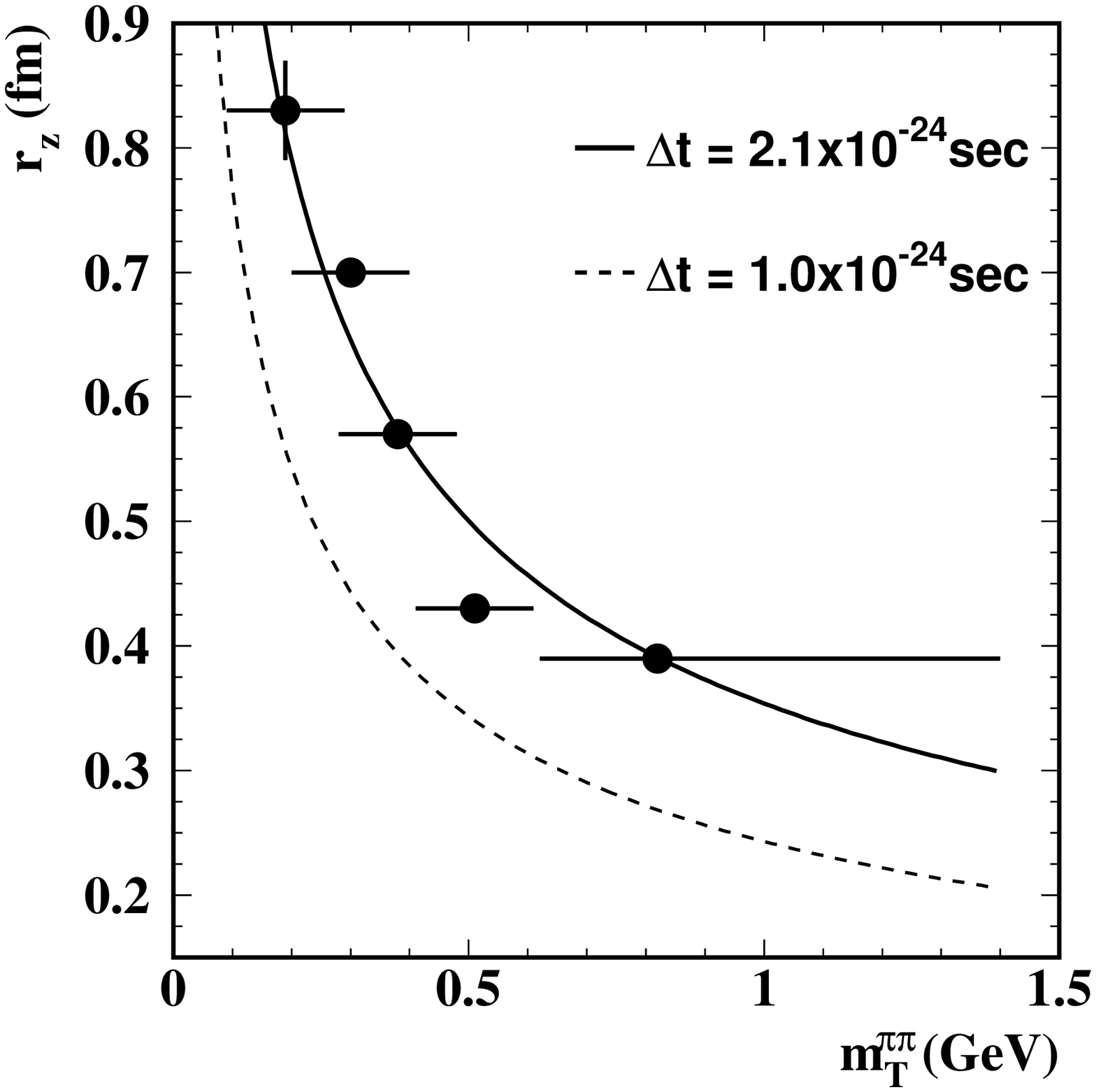}
\includegraphics[width=0.49\textwidth]{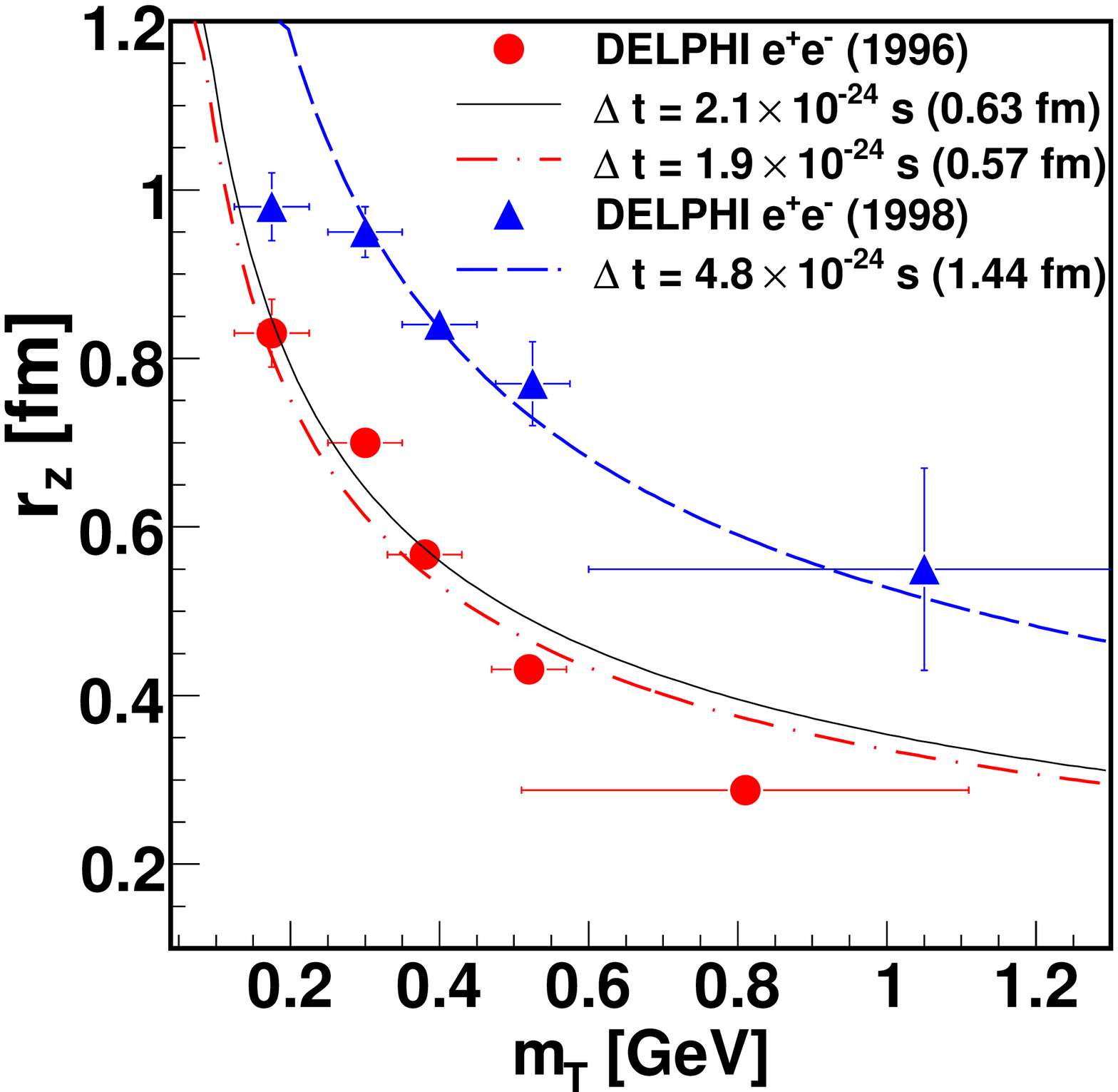}}}
\caption{The $m_T$ dependence of the longitudinal component of the HBT source size for pions; Left panel: Figure taken from~\cite{Alexander:2001gk}; Right panel: Red circles represent DELPHI data taken from~\cite{Lorstad:1996xq}. Blue triangles show data points from~\cite{Smirnova:1999aa}. See text for explanation of the curves. }
\label{fig:delphi}
\end{figure}
A very good agreement between the experimental data
and the theoretical curve is observed. Two points, however, about this figure.

Firstly, we believe that such a good agreement is partially a result of a mistake
that was made by Alexander. The author cites the following paper~\cite{Lorstad:1996xq}
as the source of the DELPHI results.
However,  we found that there is a discrepancy between the numbers presented in this paper 
and those plotted by Alexander. 
The main (and probably the only) difference seems to be in the last point 
for $m_T=0.81~GeV$ where Alexander's point is significantly
larger than DELPHI published value. The correct DELPHI data 
from~\cite{Lorstad:1996xq} has been plotted on the right panel of Figure~\ref{fig:delphi} 
represented by red circles. For comparison, we plot the prediction from
the Heisenberg uncertainty relations (Eq.~\ref{eq:HURRz} for 
$\Delta t = 2.1 \times 10^{-24}s~(0.63~fm)$ represented 
by the solid black line. This is the same curve plotted 
on the left panel of Figure~\ref{fig:delphi}.
We also performed a new fit to the correct DELPHI 
data (red circles) and found that the best $\Delta t$ that described 
the results is $1.9\times 10^{-24}s (0.57~fm)$ (dash-dotted red line).
This is not a serious mistake;
however the agreement between Eq.~\ref{eq:HURRz} and experimental data is not as good 
as Alexander showed in his paper~\cite{Alexander:2001gk}. 
Thus, we think that this problem should be mentioned here since 
the figure plotted on the left panel of Fig.~\ref{fig:delphi} 
was copiously cited in many other publications, including review
articles (e.g.~\cite{Alexander:2003ug,Kittel:2005fu}).

The second point is about the DELPHI data themselves. While Alexander refers to results 
presented in 1996 and published as proceedings in 1997~\cite{Lorstad:1996xq}  
it should be also mentioned that the same experiment published,also
as conference proceedings,  newer results   
in 1999 and presented in 1998~\cite{Smirnova:1999aa} that are different than the 
previous ones (see blue triangles on the right panel of Figure~\ref{fig:delphi}).
Presumably, the later DELPHI results represent an improvement over the earlier ones, so
it seems reasonable to use these.  Until a final result is published in a refereed journal
we have
fitted newer DELPHI data~\cite{Smirnova:1999aa} using Eq.~\ref{eq:HURRz}
and found that the best value of the $\Delta t$ that described the experimental 
data is $4.8\times 10^{-24}~(1.44~fm)$.

In heavy ion collisions, it is primarily the {\it transverse} HBT radii which reflect
collective flow.  Thus, it is interesting to try to understand the $m_T$-dependence of
femtoscopic parameters like $R_{T}$, $R_{O}$ and $R_{S}$.  Any direct connection between
these radii and the uncertainty principle is unclear.

\section{Discussion and conclusions}
\label{sec:summary}

Claims that we understand the physics of ultrarelativistic heavy ion collisions become much
more compelling when a clear comparison of data from $A+A$ and \pp collisions are made
and the differences explained.
The importance of such a comparison is obvious in studies of jet suppression from
azimuthal correlations~\cite{Adler:2002tq} and leading particle 
distributions~\cite{Adler:2003qi,Adams:2003kv}
at large $p_T$.


In this article, we presented a comprehensive review of the world systematics 
of the femtoscopic results from elementary particle collisions and 
identified trends seen in the data. 

The transverse mass dependence of the femtoscopic results from heavy ion collisions at RHIC 
is an important evidence of flow~\cite{Retiere:2003kf}. 
Surprisingly, a similar dependence is observed in small systems, 
as presented in Section~\ref{sec:smallptdep}. 
It has been even shown that not only there is the $m_T$ dependence of HBT radii
in \pp collisions at RHIC but it is very similar to what is seen in $d+Au$ and $Au+Au$ collisions suggesting 
that the only scaling between small and big system is the size~\cite{Chajecki:2005zw}.
However, it is unclear whether the  $m_T$ dependence in small systems originates also from ``flow''.
Even though such explanation has been provided by Cs{\"o}rg{\H o}
 and collaborators~\cite{Csorgo:2004id}
the literature includes rather alternative explanations.
Among them is the Heisenberg uncertainty principle
suggested by Alexander {\it et al.}~\cite{Alexander:1999rm,Alexander:2001gk}.
However, as demonstrated and discussed in Section~\ref{sec:femtosmall},
a more detailed study of the results from \epem collisions 
complicates the quantitative comparisons of the data 
and thus the interpretation.
Another physics process that could potentially generate the space-momentum correlations 
in small systems was the string fragmentation. However, as pointed out by Alexander~\cite{Alexander:2003ca}, 
the mass dependence of the experimentally observed HBT radii (especially the difference between
mesons and baryons)  is a big problem for the Lund string model.
Long-lived resonances (e.g. $\omega$) may also generate a $m_T$ 
dependence of femtoscopic radii~\cite{Wiedemann:1996ig}. However, since the resonance 
 ``halo'' length scale is fixed (e.g. $c \dot \tau_{\omega}$) 
and the ``core'' scale varies by $\approx 6x$ between \AuAu and \pp, 
one expects a different resonance-induced effect in the two systems.
Preliminary studies~\cite{Kisielprivate} with the Therminator~\cite{Kisiel:2005hn} 
model confirm this expectation.
The different effect of resonances on the transverse mass dependence of the HBT radii 
on the system size has been demonstrated by Therminator event generator~\cite{Kisielprivate}.
Finally, Bia{\l}as {\it et al.} provided a model~\cite{Bialas:2000yi}
that assumes a proportionality between the four-momentum  and the four-vector 
describing the particle space-time position at freeze-out
 and showed that it can explain the data from \epem collisions.

Other interesting observations come from  the multiplicity dependence 
of femtoscopic radii in elementary particle collisions. Surprisingly the results
from various experiments show the increase of the femtoscopic sizes with the number
of charged particles similarly as it has been observed in heavy ions collisions. 
These results should not be used as an evidence of flow but they strongly suggest
that the only difference between the size of the source in small and big systems is 
the average multiplicity. Such observation in heavy ion collisions suggests
that the system is entropy dominated~\cite{Caines:2006if}.


\end{document}